\documentclass{epl}
\usepackage{stmaryrd}
\usepackage{latexsym}
\usepackage{amssymb,amsmath}
\usepackage{graphicx,bm}

\title{Phonon-assisted resonant tunneling through a triple-quantum-dot: a phonon-signal 
detector}

\author{X. Y. Shen, Bing Dong, X. L. Lei}
\institute{Department of Physics, Shanghai Jiaotong University, 1954
Huashan Road, Shanghai 200030, China}
\date{\today}

\begin{document}

\maketitle

\begin{abstract}
We study the effect of electron-phonon interaction on current and
zero-frequency shot noise in resonant tunneling through a series
triple-quantum-dot coupling to a local phonon mode by means of a
nonperturbative mapping technique along with the Green function
formulation. By fixing the energy difference between the first two
quantum dots to be equal to phonon frequency and sweeping the level
of the third quantum dot, we find a largely enhanced current
spectrum due to phonon effect, and in particular we predict current
peaks corresponding to phonon-absorption and -emission assisted
resonant tunneling processes, which shows that this system can be
acted as a sensitive phonon-signal detector or as a cascade phonon
generator.
\end{abstract}

Phonon-assisted inelastic tunneling in semiconductor quantum dot
(QD) system at low temperature has become a focus issue in recent
years.\cite{Fujisawa,Brandes,Keil,Dong1,Bonca,Ness,Mitra,Koch,Flensberg,Galperin}
In particular, a recent experiment has measured the nonlinear
tunneling through a double-QD (DQD) with the observation of
spontaneous phonon emission leading to an additional satellite peak
in the current spectrum,\cite{Fujisawa} which can be ascribed to an
interference effect of the electron-phonon interaction (EPI) in a
DQD via nonperturbative theoretical
analyses.\cite{Brandes,Keil,Dong1} This experiment opens a
possibility of designing DQD as a coherent phonon generator.
However, the phonon-assisted peak in current spectrum is quite
fragile and thus detection of phonon-signal is a difficult task in a
DQD.\cite{Fujisawa}

In this letter, we propose a setup containing a triple QD in series
coupled to a common local phonon bath and two normal leads, in which
energy difference between the first two QDs is fixed to be equal to
the phonon frequency, i.e., head of the device acts as a phonon
emitter when there is a nonequilibrium current flowing through as
suggested by Fujisawa et al.\cite{Fujisawa} Intuitively, it is
imaginable that if the energy of the third QD is tuned, via applying
gate voltage, to be higher than the second QD with one-phonon-energy
(case a in Fig.~1 below), the emitted phonon could be re-absorbed by
electron to help electron tunneling through QD 3 resonantly,
resulting in a phonon-absorption-assisted enhanced peak in current
spectrum. That is to say that QD 3 {\em detects} the generated
phonon. On the other hand, we predict a significant enhancement of
current provided that the energy of QD 3 is further lower than QD 2
by one-phonon-energy (case c in Fig.~1), showing that {\em more}
phonon quanta are generated in tunneling process.

\begin{figure}[t]
\includegraphics [width=0.50\textwidth]{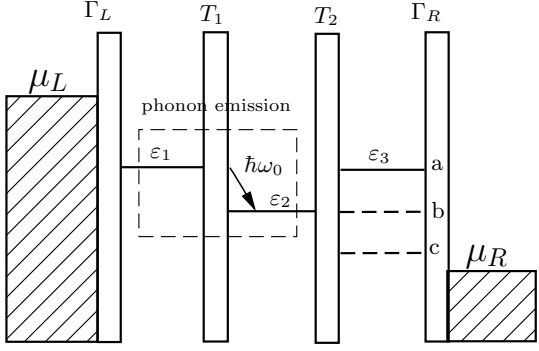}
\caption{Schematic description of tunneling of a triple-QD in the
presence of
electron-phonon interaction. $\epsilon_1-\epsilon_2=\hbar \omega_0$; a, b, and c denotes 
$\epsilon_3-\epsilon_2=\hbar \omega_0$, $0$, and $-\hbar \omega_0$,
respectively. $\mu_{L}(\mu_{R})$ responds to chemical potentials of
the lead $L$(lead $R$). $\Gamma_{L}$ and $\Gamma_{R}$ stands for the
coupling between the lead with QD. $T_{1}$ and $T_{2}$ is the
coupling between QDs.} \label{fig1}
\end{figure}

Our setup is schematically shown in Fig.~\ref{fig1}. We consider a triple QD (consisting of 
QD 1, QD 2, and QD 3) which are connected via a tunnel barrier. QD 1 and QD 3 are connected 
to an electron  reservoir in thermal equilibrium with chemical potentials $\mu_{L}$(source) 
and $\mu_{R}$(drain), respectively, with $\mu_{L}>\mu_{R}$. Its
Hamiltonian can be written as
\begin{equation}
H=H_{L}+H_{R}+H_{cen}+H_{T}, \label{ham}
\end{equation}
where $H_{L}+H_{R}$ describe the two leads, $H_{cen}$ represents the
central region consisting of a triple QD in series, and $H_{T}$ is
the coupling of the dots to the lead, respectively:
\begin{equation}
H_{L}+H_{R} =\sum_{\eta\in L,R;k}\epsilon_{\eta k}c^{\dagger}_{\eta
k}c_{\eta k},
\end{equation}
\begin{eqnarray}
\label{eq1}
H_{cen}=\sum_{\alpha}\epsilon_{\alpha}d^{\dagger}_{\alpha}d_{\alpha}+\hbar\omega_{0}
b^{\dagger}
b-\sum_{\alpha}\lambda_{\alpha}d^{\dagger}_{\alpha}d_{\alpha}(b^{\dagger}+b)
-T_{1}(d^{\dagger}_{1}d_{2}+d^{\dagger}_{2}d_{1})
-T_{2}(d^{\dagger}_{2}d_{3}+d^{\dagger}_{3}d_{2}),
\end{eqnarray}
\begin{equation}
H_{T}=\sum_{k}(V_{L}c^{\dagger}_{L k}d_{1}+V_{R}d^{\dagger}_{3}c_{R
k}+{\rm H.c.}),
\end{equation}
 where
$c^{\dagger}_{\eta k}(c_{\eta k})$ ($\eta=\{L,R\}$) and
$d^{\dagger}_{\alpha}(d_{\alpha})$ denote creation (annihilation)
operators for spinless electrons with momentum $k$ and energy
$\epsilon_{\eta k}$ in the left and right leads, and for spinless
electrons on the $\alpha$th ($\alpha=\{1,2,3\}$) QD, respectively.
$\epsilon_{\alpha}$ is the energy level of the $\alpha$th QD, and
$T_{1}(T_{2})$ stands for the interdot hopping between the QD 1 and
QD 2 (QD 2 and QD 3). $V_{L(R)}$ is the coupling constant of the
central regime with lead L(R). This sort of triple-QD system was
already realized in experiments about ten years ago, in which the
energy levels of every QD and dot-dot hoppings can all be adjusted
by applying gate voltages.\cite{Waugh} The operator $b^{\dagger}
(b)$ in Eq.~(\ref{eq1}) creates (destroys) a phonon, and
$\lambda_{\alpha}$ is the interaction constant of the electron on
the $\alpha$th dot with phonon. Here, we consider the phonon bath as
a single-mode phonon with dispersionless energy $\hbar \omega_{0}$,
and assume that phonon remains coherence during electron tunneling
processes, i.e., we take no account of phonon dissipation in the
present investigation. For simplicity, we also ignore
electron-electron interaction in this letter.

As the system under study contains a many-body problem of
electron-phonon scattering during the single-electron tunneling
process, it can not be solved analytically. Applying the
nonperturbative mapping technique suggested by Bon\v ca and
Trugman\cite{Bonca}, we can transform the many-body problem into a
multichannel one-body problem by transforming the Hamiltonian
Eq.~(\ref{ham}) in terms of a new set of electron states combined of
single-electron states and $n$-phonon states. We define a direct product states with 
single-electron states and n-phonon Fock states as:
\begin{equation}
 |\alpha,n\rangle=d_{\alpha}^{\dagger}\frac{(b^{\dagger})^{n}}{\sqrt{n!}}|0\rangle, 
\end{equation}
and
\begin{equation}
|\eta k,n\rangle=c_{\eta
k}^{\dagger}\frac{(b^{\dagger})^{n}}{\sqrt{n!}}|0\rangle,
\end{equation}
such that the electron state $|\alpha\rangle$ in QD and the state of electron in leads $|\eta 
k\rangle$ are accompanied by n phonons ($|0\rangle$ is the vacuum state). We perform this 
transformation on the electron-phonon interaction part of Eq.~(\ref{eq1}), so the many body 
problem can be mapped onto a one-body model:
\begin{equation} \sum_{\alpha
n}-\lambda_{\alpha}\sqrt{n+1}(|\alpha,n+1\rangle\langle
\alpha,n|+|\alpha,n\rangle\langle\alpha,n+1|).  \end{equation} With
defining these Dirac bracket as new operators: \begin{equation}
d_{\alpha n}^{\dagger}=|\alpha,n\rangle, c_{\eta k
n}^{\dagger}=|\eta k,n\rangle,
\end{equation}
we obtain:                                          \begin{equation}
\sum_{\alpha n}\lambda_{\alpha}\sqrt{n+1}
 (d^{\dagger}_{\alpha n+1}d_{\alpha n}+d_{\alpha n}^{\dagger}d_{\alpha
 n+1}).                                             \end{equation}
We can do the same transformation on the Hamiltonian Eq.~{\ref{ham}}
as $\widetilde{H}= \widetilde{H}_{L} +\widetilde{H}_{R} +\widetilde{H}_{cen} 
+\widetilde{H}_{T}$:\cite{Dong1}
\begin{equation}
\widetilde{H}_{L}+\tilde{H}_{R}=\sum_{\eta\in L,R;kn}\epsilon_{\eta
kn}c^{\dagger}_{\eta kn}c_{\eta kn}, \\
\end{equation}
\begin{equation}
\widetilde{H}_{cen} = \sum_{\alpha n}\epsilon_{\alpha
n}d^{\dagger}_{\alpha n}d_{\alpha n}-
 \sum_{\alpha n}\lambda_{\alpha}\sqrt{n+1}
 (d^{\dagger}_{\alpha n+1}d_{\alpha n}+{\rm
H.c.})-T_{1}(d^{\dagger}_{1n}d_{2n}+ {\rm H.c.})
-T_{2}(d^{\dagger}_{2n}d_{3n}+{\rm H.c.}),
\end{equation}
\begin{equation}
\widetilde{H}_{T}=\sum_{kn}(V_{Ln}c^{\dagger}_{L kn}d_{1n}+
V_{Rn}d^{\dagger}_{3n}c_{R kn}+{\rm H.c.}),
\end{equation}
where $d_{\alpha n}^{\dagger}$ and $ c_{\eta kn}^{\dagger}$ are new
operators with phonon quanta $n$, $\epsilon_{\alpha
n}=\epsilon_{\alpha}+n\hbar\omega_{0}$, $\epsilon_{\eta
kn}=\epsilon_{\eta k}+n\hbar\omega_{0}$. It is important to note
that the channel indices stand for the phonon quanta, so we have to
add a weight factor
$P_{n}=(1-e^{-\hbar\omega_{0}/k_{B}T})e^{-n\hbar\omega_{0}/k_{B}T}$
to the $n$th channel.

We define the retarded GFs of the triple QDs,
$G_{\alpha\beta,mn}^{r}(t,t')=-i\theta(t-t')\langle\{d_{\alpha
m}(t),d_{\beta n}^{\dagger}(t')\}\rangle$ $(\alpha,\beta=1,2,3)$,
$G_{\eta k \alpha,mn}^{r}(t,t')=-i\theta(t-t')\langle\{c_{\eta
km}(t),d_{\alpha n}^{\dagger}(t')\}\rangle$ where $m,n$ represents
the phonon quanta. Employing the equation-of-motion technique, we
get the retarded GFs in the matrix form:
\begin{eqnarray}
\begin{pmatrix}
  \bm G_{11}^{r}&\bm G_{12}^{r}&\bm G_{13}^{r}\\
  \bm G_{21}^{r}&\bm G_{22}^{r}&\bm G_{23}^{r}\\
  \bm G_{31}^{r}&\bm G_{32}^{r}&\bm G_{33}^{r}
\end{pmatrix}=
\begin{pmatrix}
 (\omega+\frac{i}{2}\Gamma_{L})\bm I-\bm A_{1}&T_{1}\bm I &0\\
 T_{1}\bm I&\omega\bm I-\bm A_{2}&T_{2}\bm I\\
 0&T_{2}\bm I&(\omega+\frac{i}{2}\Gamma_{R})\bm I-\bm A_{3}
\end{pmatrix}^{-1}, \label{gfr}
\end{eqnarray}
in which $\omega$ responds to energy, and $\bm I$ is a $N\times N$
unit matrix and $\bm A_{\alpha}$ is a $N\times N$ symmetrical
tri-diagonal matrix with $\bm
A_{\alpha;nn}=\epsilon_{\alpha}+n\hbar\omega_{0}$, $\bm
A_{\alpha;n(n-1)}=-\lambda_{\alpha}\sqrt{n}$, and $\bm
A_{\alpha;n(n+1)}=-\lambda_{\alpha}\sqrt{n+1}$, respectively.
$\Gamma_{\eta}=2\pi
\sum_{k}|V_{\eta}|^{2}\delta(\omega-\epsilon_{\eta
k}-m\hbar\omega_{0}) $ with the wide band limit represents the
coupling strength of the center region with lead $\eta$.

The current of the $n$th channel in the left lead through the center
region can be obtained from the time evolution of the occupation
number operator of the left lead:
$I=-e\langle\dot{N}_{L}\rangle=-\frac{ie}{\hbar}\langle[\tilde{H},N_{L}]\rangle$
 with $N_{L}=\sum_{k,n}c_{Lkn}^{\dagger}c_{Lkn}$. After some algebra, we find
 $I=\frac{ie}{\hbar}\sum_{k,n}[V_{Ln}\langle c_{Lkn}^{\dagger}d_{1n}\rangle-V_{Ln}^{*}\langle 
d_{1n}^{+}c_{Lkn}\rangle]$. Using the Keldysh nonequilibrium Green's
function (GF) technique and considering that the total current is a
sum over all pseudo-channels accompanied with the weight factor
$P_{n}$, we get the current as:\cite{Dong1,Bonca,Ness}
\begin{eqnarray}
\label{eq2} I=\frac{e}{h}\int d\omega \sum_{mn} (t_{Lmn}^{+}t_{Lmn})
\{ P_{n}f_{L}^{n}(\omega) [1-f_{R}^{m}(\omega)]
 -P_{m}f_{R}^{m}(\omega) [1-f_{L}^{n}(\omega)] \},
\end{eqnarray}
where $f_{L(R)}(\omega)=(1+e^{(\omega-\mu_{L(R)})/k_{B}T})^{-1} $ is
the Fermi distribution of the leads at local thermal equilibrium, in
which $T$ is the temperature. $t_{Lnm}$ represents the transmission
probability of a electron through the center region from the $n$th
channel of the left lead to the $m$th channel of the right lead.
According to the Fish-Lee relation relating the scatting matrix
elements with the retarded GFs,\cite{Fish} we describe the
transmission and reflection probabilities in terms of retarded GFs
Eq.~\eqref{gfr}:
\begin{eqnarray}
r_{L(R),mn}&=&-\delta_{mn}+i\Gamma G_{11(33),mn}^{r}(\omega),\\
t_{L(R),mn}&=&i\Gamma G_{31(13),mn}^{r}(\omega).
\end{eqnarray}
Furthermore, we use the B\"uttiker scatting method to calculate the
shot noise\cite{Buttiker} and follow our previous work,\cite{Dong1}
we obtain the zero-frequency shot noise of the system:
\begin{eqnarray}
\label{eq3}
  S_{LL}(0)&&=\frac{2e^{2}}{h}\int
  d\omega\sum_{mn}\{|(t_{L}^{\dagger}t_{L})_{mn}|^{2}P_{n}f_{L}^{n}(\omega)
   (1-f_{L}^{m}(\omega))+|(t_{L}^{\dagger}r_{L})_{mn}|^{2}[P_{n}f_{L}^{n}(\omega)\nonumber\\
  &&\times(1-f_{R}^{m}(\omega))+P_{m}f_{R}^{m}(\omega)(1-f_{L}^{n}(\omega))]
  + |(t_{R}^{\dagger}t_{R})_{mn}|^{2}P_{n}f_{R}^{n}(\omega)(1-f_{R}^{m}(\omega))\}.
\end{eqnarray}

In order to obtain a more physical view of the results, we now make
a numerical simulation of the current and the shot noise based on
Eqs.~\eqref{eq2} and \eqref{eq3} as functions of the energy
difference between QD 2 and QD 3:
$\epsilon=\epsilon_{2}-\epsilon_{3}$ (which can be tuned by applying
gate voltage). With considering the experiment by
Fujisawa,\cite{Fujisawa} we choose the temperature $T=23$mk
($k_{B}T=2\mu$eV), and dispersionless phonon $\hbar\omega_{0}=20\mu
$eV. The scale of the dot is $d=c_{s}/\omega_{0}=175nm$ where
$c_{s}=5300 m/s$ is the longitudinal sound velocity. At low
temperature the phonons are assumed to be piezoelectric acoustical
mode with interaction constant
$|\lambda|^{2}=\frac{1}{2\pi}\frac{\hbar P}{2\rho V\omega_{0}}$,
where $P$ is the piezoelectric coupling, $\rho$ is the ion mass
density, $V$ is the volume.\cite{henrik} Employing the typical GaAs
parameters,\cite{henrik} we get $\lambda\sim 1.6\mu $eV. We find the
electron-phonon interaction constant $\lambda_{1}, \lambda_{2},
\lambda_{3}$ on the three quantum dots coincide up to a phase
factor.\cite{Brandes} For simplicity, we choose phase factor is $-1$
between the QD 1 and QD 2 (QD 2 and QD 3), so we get
$\lambda_{1}=-\lambda_{2}=\lambda_{3}=1.6\mu$eV (This can be done by tuning size of each dot 
and distance between them in experiments). In this cases, we set the
couplings of the dots with the two leads to be symmetric,
$\Gamma_{L}=\Gamma_{R}=\Gamma=4\mu$eV, and the interdot hoppings to
be equal, $T_{1}=T_{2}=T_{c}=4\mu$eV.  We assume
$\epsilon_{2}=0\mu$eV, which is equal to the Fermi energies of the
left lead and right lead at equilibrium. We also set the energy gap
of the QD 1 and QD 2 as $\epsilon_{1}-\epsilon_{2}=\hbar
\omega_{0}=20\mu$eV and the symmetrically applied bias voltage
$\mu_{L}-\mu_{R}=eV_{sd}=200\mu $eV so that
$\mu_{L}\gg\epsilon_{1},\epsilon_{2},\epsilon_{3}\gg\mu_{R}$, in
which condition sufficient more pseudo-channels are involved in
transport and thus our numerical results are nearly independent of
the bias-voltage.

\begin{figure}[t]
\includegraphics [width=0.5\textwidth]{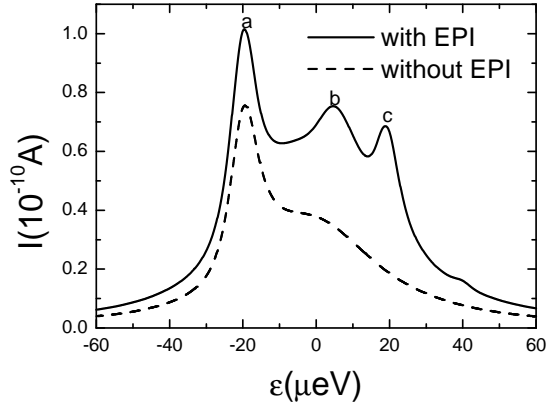}
\caption{Calculated total tunneling current versus energy difference
$\epsilon$ with parameter: $\Gamma=4\mu$eV and
$\lambda_{1}=-\lambda_{2}=\lambda_{3}=1.6\mu$eV, the temperature of
the system $T=23$mK.} \label{fig2}
\end{figure}

In Fig.\ref{fig2} we plot the total current of the triple-QD system,
for comparison, in the absence of EPI (dash line) vs the energy
difference $\epsilon$. It is interesting to find a peak at
$\epsilon=\epsilon_2-\epsilon_3=-\hbar \omega_0$ (where
$\epsilon_{1}=\epsilon_{3}$) in the system without considering
electron-phonon interaction. This peak is due to a resonant
tunneling between QD1 and QD3 via QD 2. We can describe the dynamics
of the triple QD system with quantum rate equations for the density
matrix of the dots.\cite{Renzoni} With solving the master equation
algebraically, we get the current at zero temperature analytically
as:
\begin{equation}
I=\frac{e}{\hbar} \frac{4\Gamma
T_{c}^{4}}{16T_{c}^{4}+(2\Gamma^{2}+8\epsilon^{2}+16\epsilon\epsilon_{12}+8\epsilon_{12}^{2})
T_{c}^{2}+4\epsilon_{12}^{4}+\epsilon_{12}^{2}\Gamma^{2}+4\epsilon_{12}^{2}\epsilon^{2}+ 
8\epsilon_{12}^{3}\epsilon},
\end{equation}
with $\epsilon_{12}=\epsilon_{1}-\epsilon_{2}$. It is obvious to
obtain the peak of current at
$\epsilon=-\epsilon_{12}=\hbar\omega_{0}$.

In Fig.\ref{fig2} we also plot the total current of the triple-QD
system in the presence of EPI vs the energy difference $\epsilon$.
Similar with the current spectrum of a double-QD
system,\cite{Fujisawa,Brandes,Keil,Dong1} an obvious overall
enhancement of current is found for the triple-QD system with phonon
bath in comparison with that without EPI. Moreover, as above
expected, our results predict three current peaks labeled by a, b,
and c, which are corresponding to the three configurations described
in Fig.~1. When an extremely large bias-voltage is applied to the
whole system, an electron can resonantly jump from QD 1 to QD 2
accompanied by one-phonon generated, then if the system is at
configuration b, $\epsilon_2=\epsilon_3$, the electron can directly
resonantly tunnel through the device and the current peak b is
caused by phonon emission between QD1 and QD2; however, if the
system is at configuration c, $\epsilon_2-\epsilon_3=\hbar\omega_0$,
the electron will emit another phonon to reach QD 3, leading to
two-phonon-emission peak c; more interestingly, if the system is at
configuration a, $\epsilon_2-\epsilon_3=-\hbar \omega_0$, an
enhanced current peak a also occurs stemming from re-absorption of
the generated phonon in head of the device by electron to overcome
the barrier between QD 2 and QD 3, which is enhanced by $30$\% in comparison to the resonant 
peak without EPI. Of course, phonon dissipation to environment could lessen these 
phonon-assisted resonant peaks. Albeit we do not consider
nonequilibrium phonon effect\cite{Mitra,Koch} and dissipation due to
environment,\cite{Flensberg,Galperin} however, it can be expected from our above calculations 
that the multi-QD system could function as a cascade phonon
generator with higher emission efficiency than a double-QD or as a
sensitive and
``good" phonon-signal detector. Actually, we can put this system into a phonon resonant 
cavity in experiments to lessen the leakage of phonon and thus to enhance the probability of 
re-absorption of phonon. These studies will be the subject in our
future work.

\begin{figure}[t]
\includegraphics [width=0.8\textwidth]{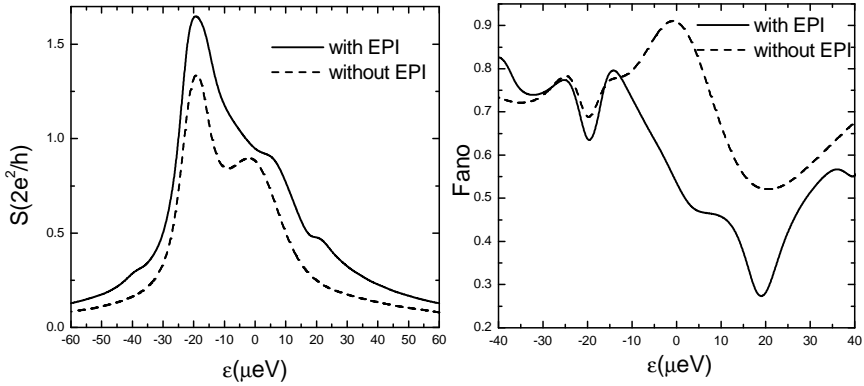}
\caption{(a) Zero-frequency shot noise versus energy difference
$\epsilon$ with the same parameter of Fig.\ref{fig2}; (b) Fano factor versus energy  
difference $\epsilon$.} \label{fig3}
\end{figure}

We also examine the zero-frequency shot-noise (a) and Fano factor $F=S_{LL}/2eI$ (b)
 with the energy difference $\epsilon$, as plotted in Fig.\ref{fig3}. Corresponding to the 
three current peaks
due to phonon-absorption or -emission, the shot noise spectrum shows
also single peak in the phonon-assisted resonant tunneling region.
As the shot noise $S\sim T_{tr}(1-T_{tr})$, which $T_{tr}$ is
transmission probability, there is no shot noise generation when the
$T_{tr}=1$ or $0$. The maximal generation of the shot noise occurs
while the transmission probability is between $0$ and $1$. We also
find a small peak at $\epsilon=0$ in the absence of phonon. However,
the shoulder at $\epsilon=0$ is smeared smoothly by the interaction
of phonon. The Fano factor $F$ displays three dips at the three
points a, b, and c, implying that the inelastic resonance suppresses
the shot noise. It is worth noticing that the phonon-absorption-assisted tunneling induces a 
more pronounced dip.

In summary, we investigate the resonant tunneling through a triple
QD in the presence of EPI at low temperature by means of the
nonperturbative mapping technique in combination with the
nonequilibrium GF method. By making the first two dots act as a
phonon emitter and sweeping the energy level of the third QD, we
evaluate the tunneling current and predict that resonant peaks in
the current spectrum are not only due to spontaneous phonon-emission
assisted tunneling, but also due to phonon-absorption process. We
also study zero-frequency shot noise of this sytem.

This work was supported by Projects of the National Science
Foundation of China, the Shanghai Municipal Commission of Science
and Technology, the Shanghai Pujiang Program, and NCET.

\end{document}